\documentclass[a4paper]{spie}  %>>> use this instead for A4 paper
\usepackage[]{graphicx}

\title{High spatial resolution and high contrast optical speckle imaging with FASTCAM at the ORM}

\author{Lucas Labadie\supit{a,b}, 
Rafael Rebolo\supit{a,f}, 
Bruno Femen\'ia\supit{a,b}, 
Isidro Vill\'o\supit{c}, 
Anastasio D\'iaz-Sanchez\supit{c}, 
Alejandro Oscoz\supit{a}, 
Roberto L\'opez\supit{a}, 
Jorge A. P\'erez-Prieto\supit{a}, 
Antonio P\'erez-Garrido\supit{c}, 
Sergi R. Hildebrandt\supit{d}, 
Victor B\'ejar-S\'anchez\supit{a,b}, 
Juan Jos\'e Piqueras\supit{e}, 
Luis Fernando Rodr\'iguez\supit{a}
\skiplinehalf
\supit{a} Instituto de Astrofisica de Canarias, C/ Via Lactea s/n, La Laguna, Tenerife E-38200, Spain \\
\supit{b} Departamento de Astrofisica, Universidad de La Laguna, 38205 La Laguna, Tenerife, Islas Canarias, Spain \\
\supit{c} Universidad Politecnica de Cartagena, Campus Muralla del Mar, Cartagena, Murcia E-30202, Spain \\
\supit{d} Laboratoire de Physique Subatomique et de Cosmologie, 53 Avenue des Martyrs, 38026 Grenoble, France \\
\supit{e} Max-Planck-Institut f\"ur Sonnensystemforschung, Max-Planck-Str. 2, 37191 Katlenburg-Lindau, Germany \\
\supit{f} Consejo Superior de Investigaciones Cientificas, Spain \\
}

\authorinfo{Further author information: (Send correspondence to L.Labadie)\\L.Labadie: E-mail: labadie@iac.es, Telephone: +34 922 605 752
}
\begin{document} 
  \maketitle 

%%%%%%%%%%%%%%%%%%%%%%%%%%%%%%%%%%%%%%%%%%%%%%%%%%%%%%%%%%%%% 
\begin{abstract}
In this paper, we present an original observational approach, which combines, for the first time, traditional speckle imaging with image post-processing to obtain in the optical domain diffraction-limited images with high contrast (10$^{-5}$) within 0.5 to 2 arcseconds around a bright star. The post-processing step is based on wavelet filtering an has analogy with edge enhancement and high-pass filtering.  Our I-band on-sky results with the 2.5-m Nordic Telescope (NOT) and the lucky imaging instrument FASTCAM show that we are able to detect L-type brown dwarf companions around a solar-type star with a contrast $\Delta$I$\sim$12 at 2$^{\prime\prime}$ and with no use of any coronographic capability, which greatly simplifies the instrumental and hardware approach.  This object has been detected from the ground in J and H bands so far only with AO-assisted 8-10 m class telescopes (Gemini, Keck), although more recently detected with small-class telescopes in the K band. Discussing the advantage and disadvantage of the optical regime for the detection of faint intrinsic fluxes close to bright stars, we develop some perspectives for other fields, including the study of dense cores in globular clusters. % possible detection of the reflected light by planetary-mass companions.
To the best of our knowledge this is the first time that high contrast considerations are included in optical speckle imaging approach.
\end{abstract}

%>>>> Include a list of keywords after the abstract 


%%%%%%%%%%%%%%%%%%%%%%%%%%%%%%%%%%%%%%%%%%%%%%%%%%%%%%%%%%%%%
\section{INTRODUCTION}

To date, a high number of ambitious scientific cases of modern galactic and extragalactic astronomy rely on the need of very deep and sharp images able to unveil faint details. Such goals stress the importance of observations with high sensitivity, high dynamic range and high spatial resolution. In the last ten to twenty years, these requirements have been fulfilled with the development of a new class of large 8-10\,m telescopes equipped with adaptive-optics systems to compensate for the atmospheric turbulence that strongly degrades the observational needs mentioned above. Space-based observatories such as the 2.5-m HST -- or the future JWST -- have been so far the only alternative to AO-based instruments to obtain highly detailed images.\\
In the field of low-mass stars, brown dwarfs and exoplanets, the potential of this approach is well recognized and extremely vast: direct imaging at high spatial resolution and high contrast permits to reveal multiple systems and planet candidates with longer periods ($\sim$0.1$^{\prime\prime}$) than what typically probed with radial velocities, to characterize the circumstellar environments of young forming planetary systems \cite{Bally2000}, or to determine unambiguously dynamical masses by following over few years the orbit of close brown dwarfs binaries in order to calibrate the mass-luminosity relationship for substellar objects \cite{Bouy2004,Zapatero2004}. In other words, the determination of masses of very low-mass and planetary mass objects, as well as the establishment of multiplicity statistics and orbital properties are among the most fundamental parameters that can be measured experimentally using the ``direct imaging'' approach, which in return will help us to constrain star and planet formation theories.\\
From the ground, high angular resolution (HAR) observations rely on adaptive-optics systems, which are primarily designed to operate in the near-infrared regime (J, H, K bands). In the optical regime ($\lambda$$<$1\,$\mu$m), complementary data to the infrared ones are necessary for a full characterization of the spectral energy distribution, key to the determination of effective temperatures and bolometric luminosities. However, one is generally limited at these wavelengths by the poor correction of conventional near-IR optimized AO systems, unable to deliver diffraction limited resolution over significant ($>$5$^{\prime\prime}$) field of view. \\
Optical speckle imaging techniques \cite{Mackay2004,Law2006} share similar goals with near-IR AO-based instruments. The strong advent of fast readout and low readout-noise CCD has favoured the development of ``Lucky Imaging'' instruments able to deliver optical diffraction-limited images in which faint companions and close binaries can be detected at visible wavelengths, offering complementary scientific outcomes to near-IR observations. In this paper, we present new results obtained with the FASTCAM instrument \cite{Oscoz2008} developed at the Instituto de Astrofisica de Canarias which address both high angular resolution {\it and} high dynamic range issues. % in the context of very low-mass objects. 
After a short presentation of the instrument, we give an overview of its scientific capabilities, stressing on the potential for high dynamic range imaging in association with image post-processing.
%---------------%------------------
%---------------%------------------
% - the potential for other science
%---------------%------------------
%---------------%------------------

% for the determination of orbits, dynamical masses in multiple systems, multiplicity statistics, study of the circumstellar environment to constrain star and planet formation theories. For instance, the measurements of model-independent dynamical masses in close binary systems are highly requested to calibrate the mass-luminosity relationship of substellar objects. This can be done by following over few years the orbit to determine the orbital parameters, and finally the individual masses.
% Direct imaging at high spatial resolution and high contrast permits to reveal multiple systems candidate with longer periods (~0.1$^{\prime\prime}$) than what probed with radial velocities sensitive to shorter periods. 
% the calibration of the mass-luminosity relationship of substellar objects can be sought 
% The field of low-mass stars and exoplanets is particularly interested in this ... 

\begin{figure}[b]
\centering
%\begin{minipage}[t]{\textwidth}
\includegraphics[width=13.0cm]{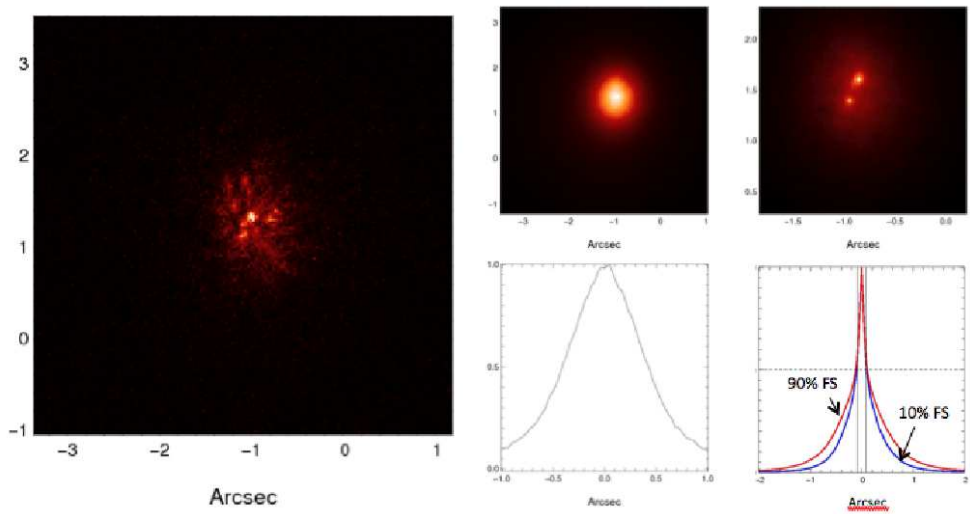}
%\end{minipage}
\caption{{\it Left:} short-exposure image of the binary star GN Tau. The pattern is composed of different speckles with different brightness. {\it Center:} The seeing limited image of the source after a long exposure. The various speckles add randomly, resulting in an diameter independent FWHM of the PSF. {\it Right:} The shift-and-add plus frame selection image.}\label{image}
\end{figure}

\section{Instrument Description}

\vspace{0.5cm}

\subsection{Exploiting diffraction-limited speckles}

As soon as the size of the telescope overcomes the Fried parameter $r_0$ that describes the atmospheric turbulence at a given wavelength, a point-like source image will appear as a rapidly spatially and temporally varying {\it speckle} pattern within the characteristic timescale of the turbulence. In long-exposure images, this effect will smear out the diffraction-limited point spread function (PSF), leading to {\it seeing-limited} images for which the spatial resolution becomes independent of the size of the telescope (see Fig.~\ref{image}). \\
\\
{\it Speckle imaging} has been one of the pioneering high resolution imaging technique able to take advantage from the high spatial resolution information contained in speckles, each of them representing an image of the source at the diffraction limit of the telescope. Speckle imaging is not a new technique: it has been implemented since the seventies by different groups \cite{Labeyrie1970,Weigelt1977}, but it is only in the last years that {\it optical} speckle imaging made real advance in astronomy (see McCay et al., these proceedings) essentially thanks to the very strong improvement of optical CCD detectors. Indeed, because this technique relies on the acquisition of short-exposure images, the readout-noise of the detector becomes the dominant noise source limitation. Going towards fainter sources requires to decrease as much as possible the readout noise, which becomes now possible with the new generation of fast readout CCD detectors.\\
In brief, speckle imaging stores separated short-exposure images lasting few tens to few hundred of milliseconds, each of them presenting a random realization of the atmospheric turbulence. 
%: in such a pattern, each speckle represents the diffraction-limited image given by the telescope of the astrophysical source observed. 
The next step is to recenter all the images of the dataset along the brightest speckle -- determined by some metrics, e.g. the brightest pixel in a given region -- using a {\it shift-and-add} technique to collapse the image cubes. The final PSF is composed of a diffraction-limited core resulting from the co-addition of the brightest speckles in each frame, surrounded by a halo -- the ``seeing disk'' -- caused by the superposition of the other speckles in the image, and similarly present in classical AO images due to imperfect correction of the turbulent wavefronts (see Fig.~\ref{image}). \\
By applying a selection of a given percent of best frames in the sequence, it is possible to improve at the same time the spatial resolution (decreasing the FWHM of the PSF) and reduce the relative contribution of the halo. This is because the best frames in the sequence corresponds to the lowest turbulence effect of the atmosphere (i.e. larger r$_{\rm 0}$). This is illustrated in the right side of Fig.~\ref{image}. ``Best frames" is intended here as those frames where most of the intensity in the image is concentrated in one given speckle.

\subsection{The optical speckle camera FASTCAM}

\begin{figure}[b]
\centering
%\begin{minipage}[t]{\textwidth}
\includegraphics[width=8.0cm]{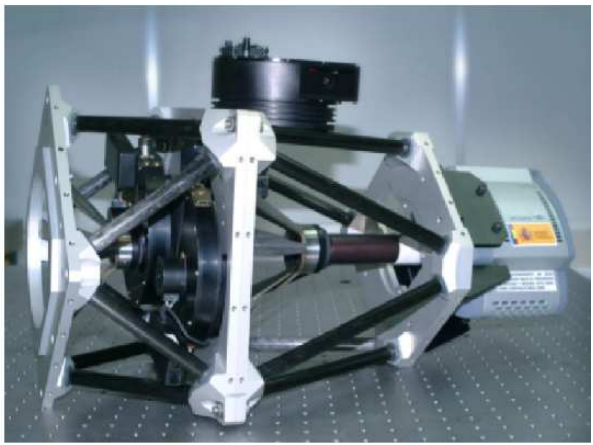}
%\end{minipage}
\caption{View of the FASTCAM camera on its mechanical support in the standard common-user instrument configuration. The speckle camera is seen on the right part of the image. The mechanical support holds also two filter wheels.}\label{fastcam}
\end{figure}

FASTCAM is a collaborative project developed by different teams at the Instituto de Astrofisica de Canarias, Canary Islands and at the Universidad Politecnica de Cartagena, in Spain. The current view of the instrument after the opto-mechanical design phase is shown in Fig.~\ref{fastcam}. The FASTCAM instrument has a great versatility and can equip different telescopes of the ORM observatory\,\footnote{ORM: Observatory of Roque de los Muchachos} such as the 2.5-m Nordic Telescope (NOT)\,\footnote{www.not.iac.es} or the 4.2-m William Herschel Telescope (WHT)\,\footnote{www.ing.iac.es}. Most of the time, FASTCAM is installed as a common-user instrument at the 1.5-m Carlos Sanchez telescope (TCS). 
\newline
\newline
Our speckle camera is based on a commercial fast readout (EM)CCD detector developed by Andor Technology. Basically, this new generation of optical detectors present the strong advantage of performing the ``Electron multiplication'' (EM) of the signal before the readout of the detector, which means that, proportionally, the readout noise can be significantly reduced in each pixel compared to useful signal, down to $\sim$1\,e- rms. The current detector is 512$\times$512 pixels, with possibility of windowing the array to speed up the readout. The individual integration time can be effectively set down to 30\,ms. The quantum efficiency is quite high, peaking at 95\% at 0.55\,$\mu$m.
\newline
\newline
The optical system of FASTCAM is based on relatively simple doublet lenses that permit to deliver various pixel scales, namely 70 to 50\,mas/pix at the TCS, 30\,mas/pix at the NOT and 13.5\,mas/pix at the WHT. Currently, no atmospheric dispersion compensator (ADC) is used with FASTCAM, although a dedicated ADC is being designed. However, this issue is not very constraining as long as we observe as low airmasses or in narrow band filters. A motorized filter wheel permits to install the majority of the wide band and narrow band optical filters typically used for imaging at visible wavelengths.

\begin{figure}[h]
\centering
%\begin{minipage}[t]{\textwidth}
\includegraphics[width=12.0cm]{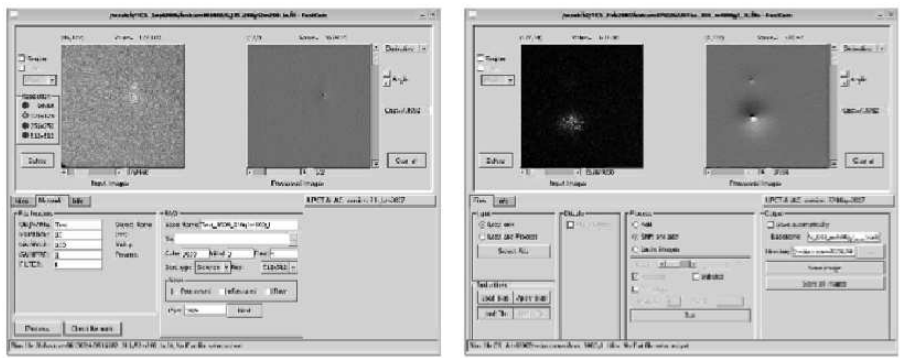}
%\end{minipage}
\caption{Screen shot of the computer interface controlling Fastcam (left) and the data reduction software (right) developed by the University of Cartagena, Spain.}\label{fastcam}
\end{figure}

\noindent FASTCAM benefits from its own software interface to control the instrument at the telescope, as well as from a dedicated data reduction software used for both an accurate data reduction of the speckle images and for a quick view of the data at the telescope. The control software and data reduction pipeline were developed at the University of Cartagena.

\section{Science with FASTCAM}

\subsection{Multiple systems at high angular resolution}

\begin{figure}[h]
\centering
%\begin{minipage}[t]{\textwidth}
\includegraphics[width=10.0cm]{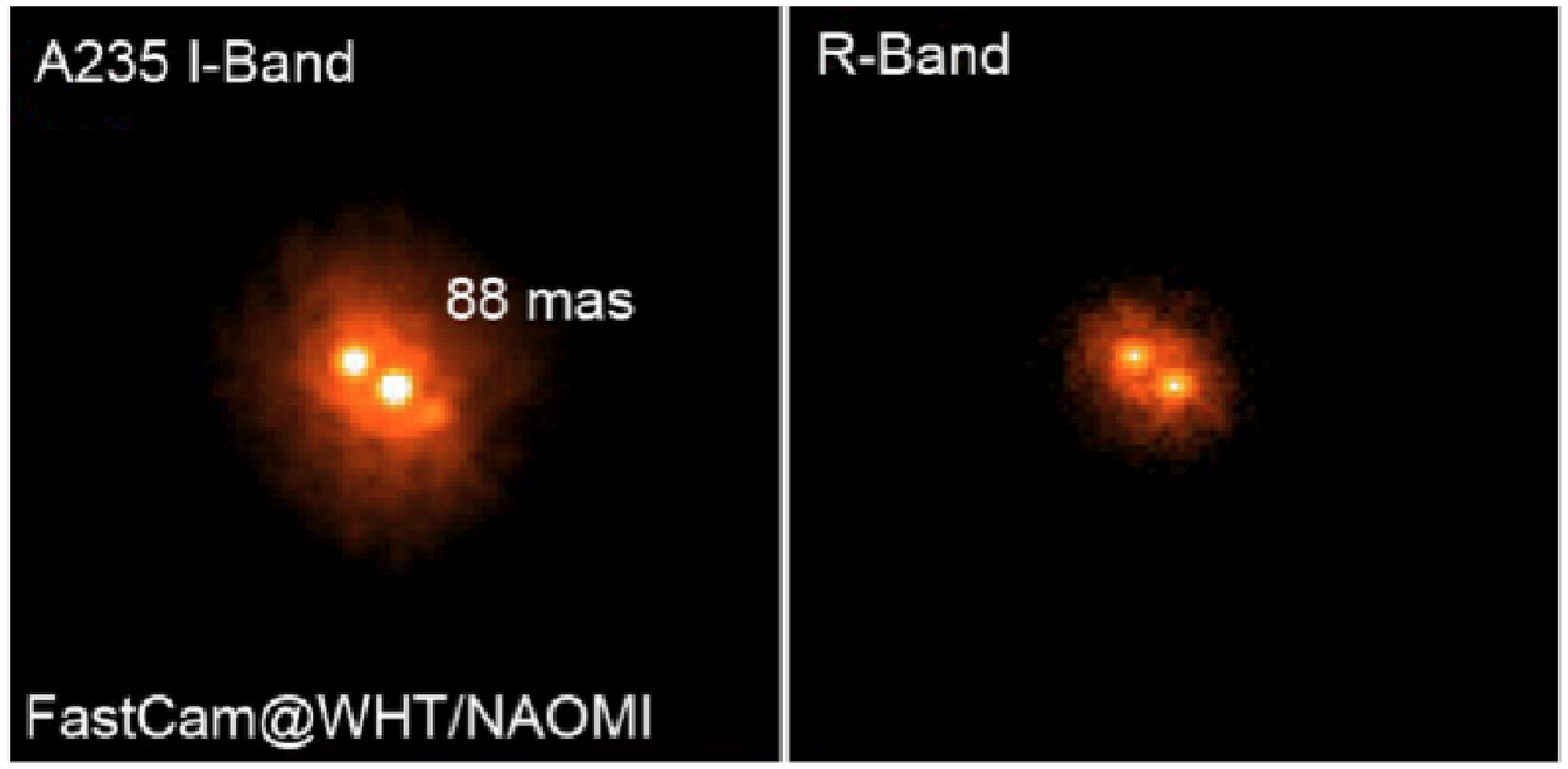}
\hspace{1.0cm}
\includegraphics[width=5.0cm]{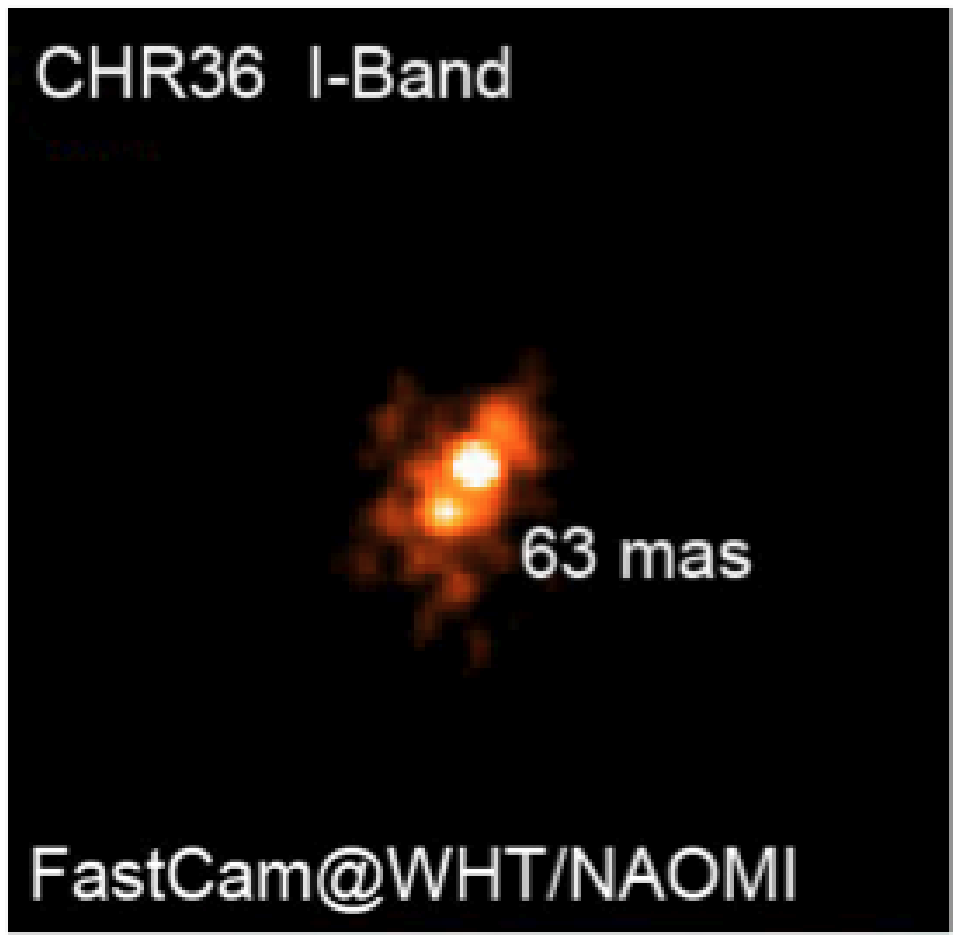}
%\end{minipage}
\caption{Close binaries images in the I-band and R-band using FASTCAM at the 4.2-m William Herschel Telescope.}\label{binary}
\end{figure}

Speckle imaging has been primarily used for imaging close sub-arcsecond multiple systems from the ground in the optical, while they were generally accessible only using AO systems in the near-infrared. Monitoring of binary stars is one of the main objective of FASTCAM, in particular those with small angular separation and relatively close in order to determine in a reasonable time scale the orbital parameters of the binary, which ultimately provide dynamical masses of the system.\\
Different observing campaigns have been conducted at the three different telescopes, reaching close to diffraction-limited images in each occasion. Fig.~\ref{binary} illustrates some of the results obtained with FASTCAM in $R$ and $I$ bands at the 4.2-m WHT. For instance, the binary system CHR\,36 with a separation of 0.063$^{\prime\prime}$ was resolved in the $I$-band filter centered at 790\,nm using integration times of the order of 50\,ms.

\begin{figure}[h]
\centering
%\begin{minipage}[t]{\textwidth}
\includegraphics[width=10.0cm]{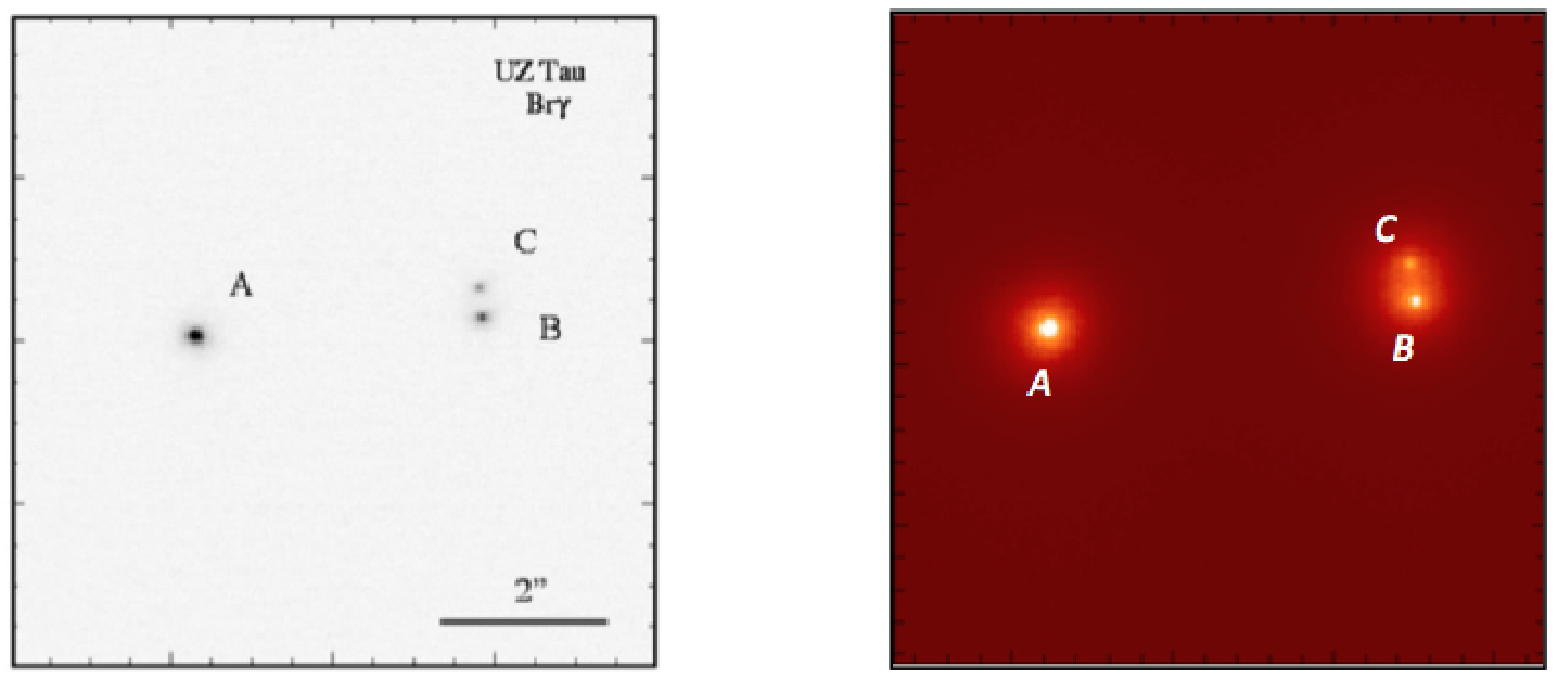}
\hspace{2.0cm}
\includegraphics[width=4.22cm]{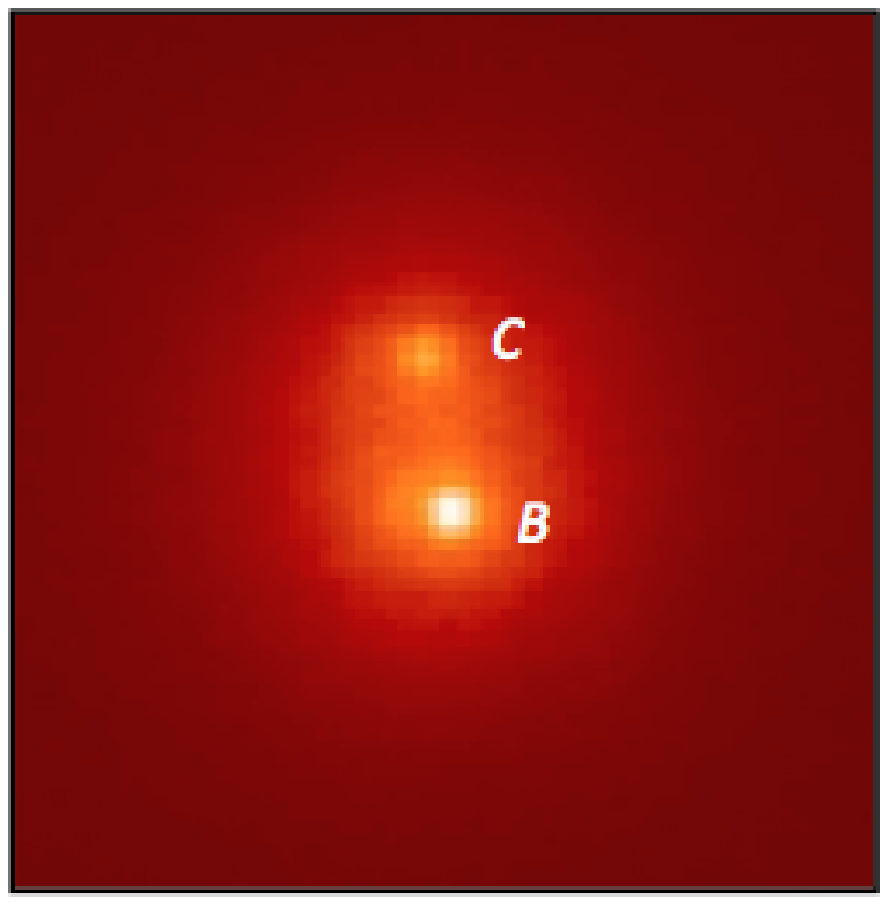}
%\end{minipage}
\caption{Comparison of infrared and optical images of the triple system UZ Tau obtained with NACO (left) and FASTCAM at the NOT (right).}\label{UZTau}
\end{figure}

\noindent Speckle imaging is not only limited to a small field-of-view in which to observe binaries. High resolution imaging can be performed on a larger field-of-view, which can be illustrated with the notion of isoplanatic angle. The frame selection process keeps those ``best'' frames corresponding to the lowest atmospheric turbulence (i.e. larger r$_{\rm 0}$) for which the isoplanatic angle is larger than the average one obtained when 100\,\% of the frames are maintained. Hence, the finest frame selection can guarantee isoplanatic angles of 30$^{\prime\prime}$ or larger, significantly larger than the average isoplanatic angle of $\sim$3\,$^{\prime\prime}$ in normal seeing conditions. This can be successfully exploited for imaging crowded fields in dense cores.\\
\\
Fig.~\ref{UZTau} compares two images of the triple system UZ\,Tau with comparable resolution. In the center part, UZ Tau is imaged at 2.16\,$\mu$m with the AO-equipped NACO instrument at the VLT \cite{Correia2006}. On the right is shown the optical image counterpart obtained at 0.8\,$\mu$m with FASTCAM at the 2.5-m NOT. The relative brightness of each component at 0.8\,$\mu$m is similar to the infrared case, which provides indications on the spectral type of the system components. The AB separation is $\sim$3.5$^{\prime\prime}$ while the BC separation is $\sim$0.36$^{\prime\prime}$.

\subsection{The question of high dynamic range at short angular separations}

\begin{figure}[h]
\centering
%\begin{minipage}[t]{\textwidth}
\includegraphics[width=3.24cm]{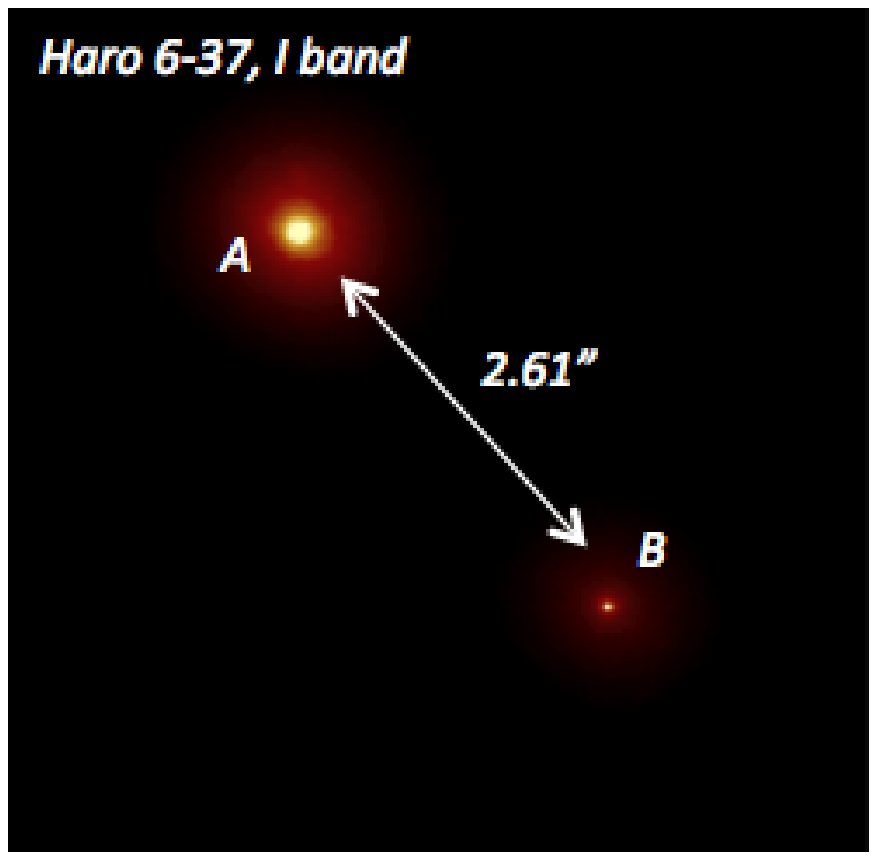}
%\hspace{0.1cm}
\includegraphics[width=3.3cm]{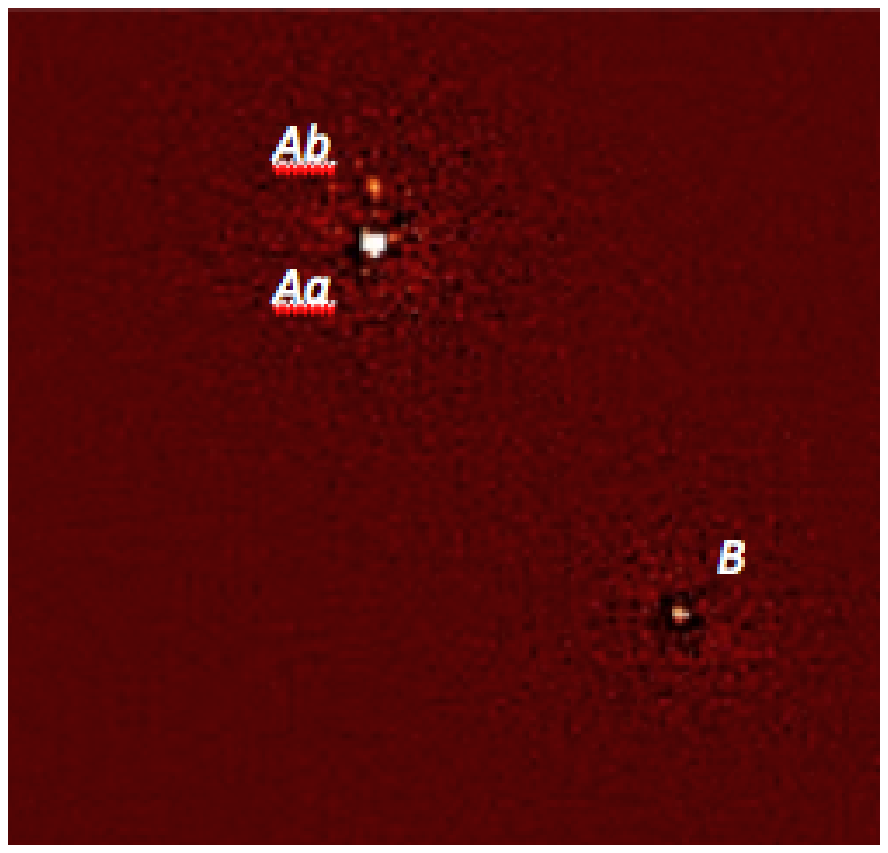}
%\hspace{0.1cm}
\includegraphics[width=3.3cm]{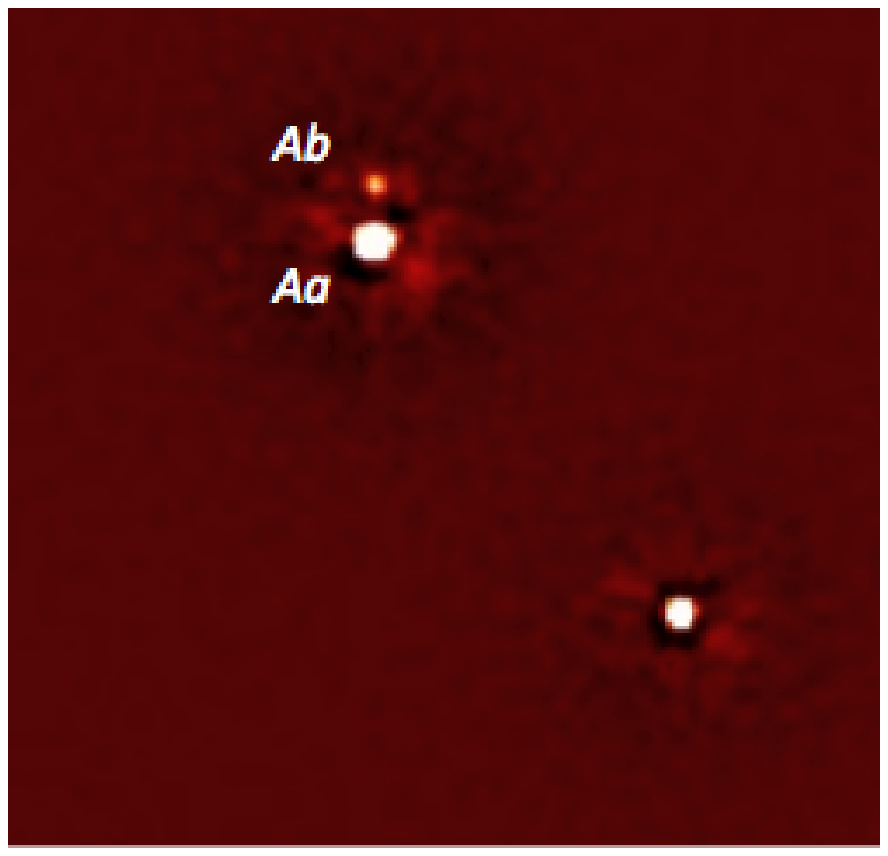}
%\hspace{0.1cm}
\includegraphics[width=3.3cm]{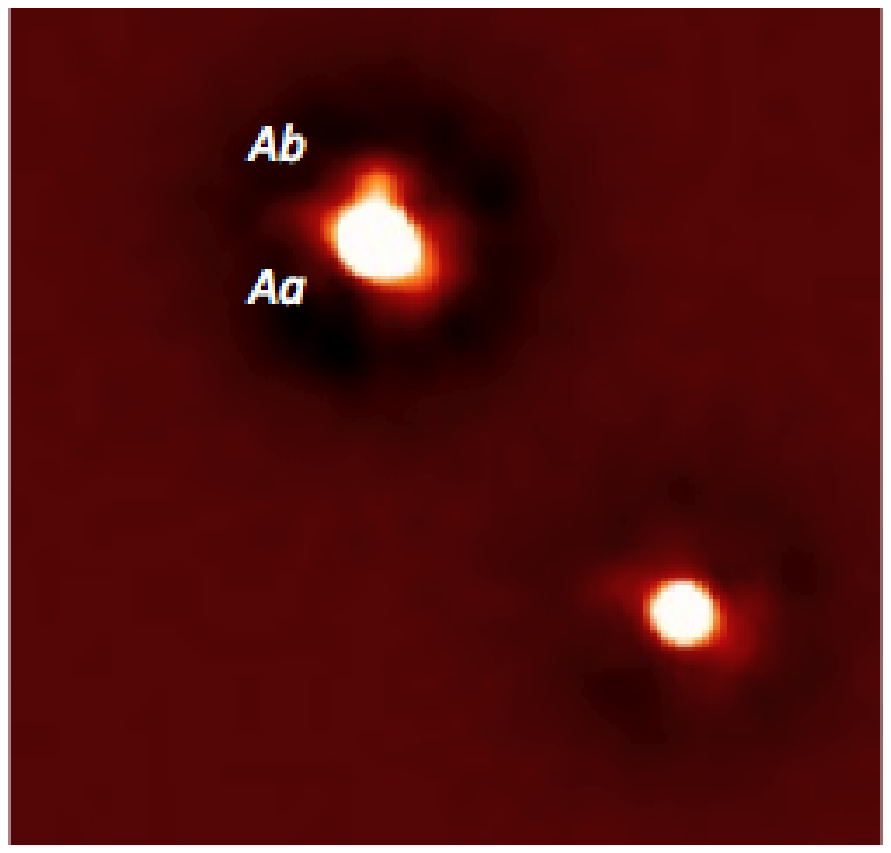}
%\hspace{0.1cm}
\includegraphics[width=3.3cm]{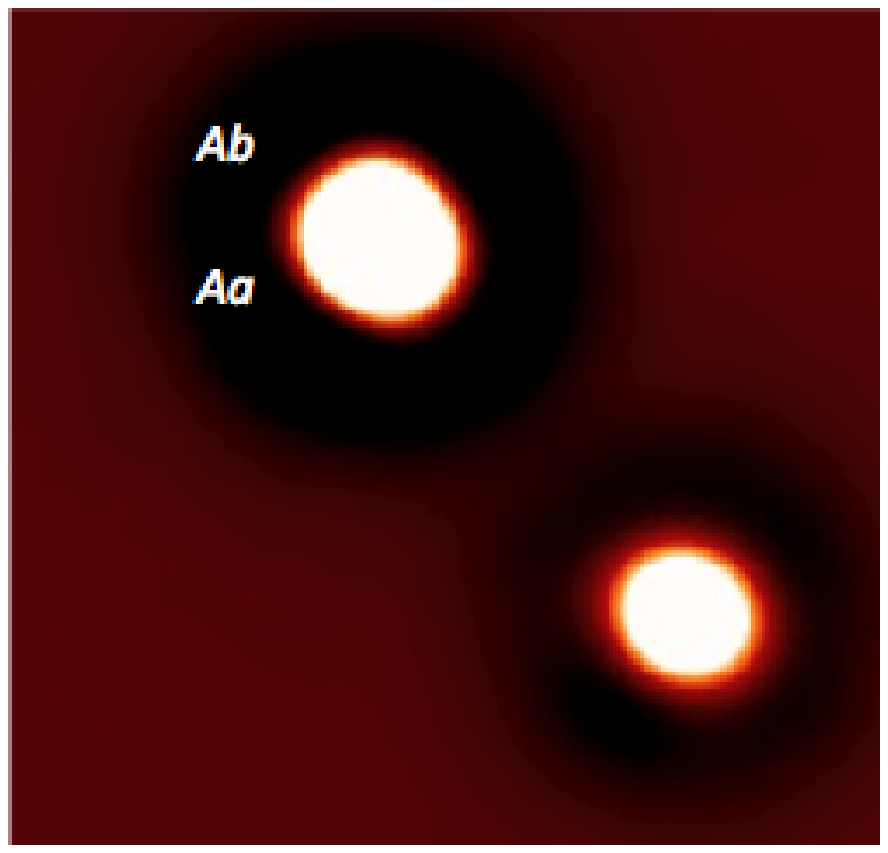}
%\end{minipage}
\caption{Wavelet decomposition of the lucky image of the system Haro\,6-37 observed with the NOT. See text for details.}\label{wavelet}
\end{figure}

On the right side of Fig.~\ref{UZTau} is shown a close view of the resolved system UZ\,Tau\,BC, which at the same time highlights a critical issue of lucky imaging: around each of the two components the seeing halo is likely masking any faint structure or companion that may lie at short angular separations. For classical high contrast long-exposure imaging, this question is traditionally solved with the use of a coronograph placed in the intermediate image plane that strongly reduces the flux contribution of the star on-axis and consequently of the halo residual left by the AO. In absence of any coronograph, a possible approach with speckle imaging is to reach as much as possible the true diffraction-limited case that corresponds to the highest Strehl ratios. This can be done with speckle {\it image reconstruction} for which several methods have been developed in the past, such as bi-speckle interferometry, speckle holography, blind deconvolution etc...\\
A second approach, that we decided to adopt in this study, is based on numerical post-processing of the lucky image in order to filter out low spatial frequencies and enhance other ones, in particular those related to faint nearby companions. This approach is based on the wavelet decomposition of the image -- a standard and routinely used image processing technique in computer sciences -- that permits to identify the best trade-off between the optimal spatial frequency at which the companion is revealed and the accessible signal-to-noise. Here, we implemented an iterative wavelet filtering described by the relationship $I$=$I$-$I$$\ast$$k$, where $k$ is the size variable wavelet kernel. The wavelet filtering process is illustrated in the sequence of Fig.~\ref{wavelet}. The first image shows the FASTCAM/NOT image at 0.8\,$\mu$m of the triple system Haro\,6-37. The A component is actually a double system with a fainter companion Ab at 0.36$^{\prime\prime}$ from Aa. In the direct lucky image, this companion can be hardly detected in the PSF halo surrounding Aa. The four successive images show the multi-resolution decomposition of the first image, with the kernel size changing by a factor 2 from the second to the fifth image. By comparing the spatial scale at which Ab is detected and the visual SNR of the detection, it appears that the second image of the sequence is the optimal trade-off between those two factors, while in the two last cases we can see that the spatial scale at which the image is filtered results in the partial or complete removal of the companion. The rebound dark zone surrounding the various components is an intrinsic feature of the filtering process.

\vspace{0.25cm}

\noindent In practice, the wavelet decomposition is substituted with wavelet {\it filtering} in which only the optimal component of the decomposition is maintained as the final post-processed image. In such a way, we benefit from a relatively simple filtering algorithm to enhance the detection of faint and nearby point-like sources. The effect of the wavelet post-processing on faint extended source is not treated here, although it certainly present a strong interest.

\begin{figure}[b]
\centering
%\begin{minipage}[t]{\textwidth}
\includegraphics[width=4.5cm]{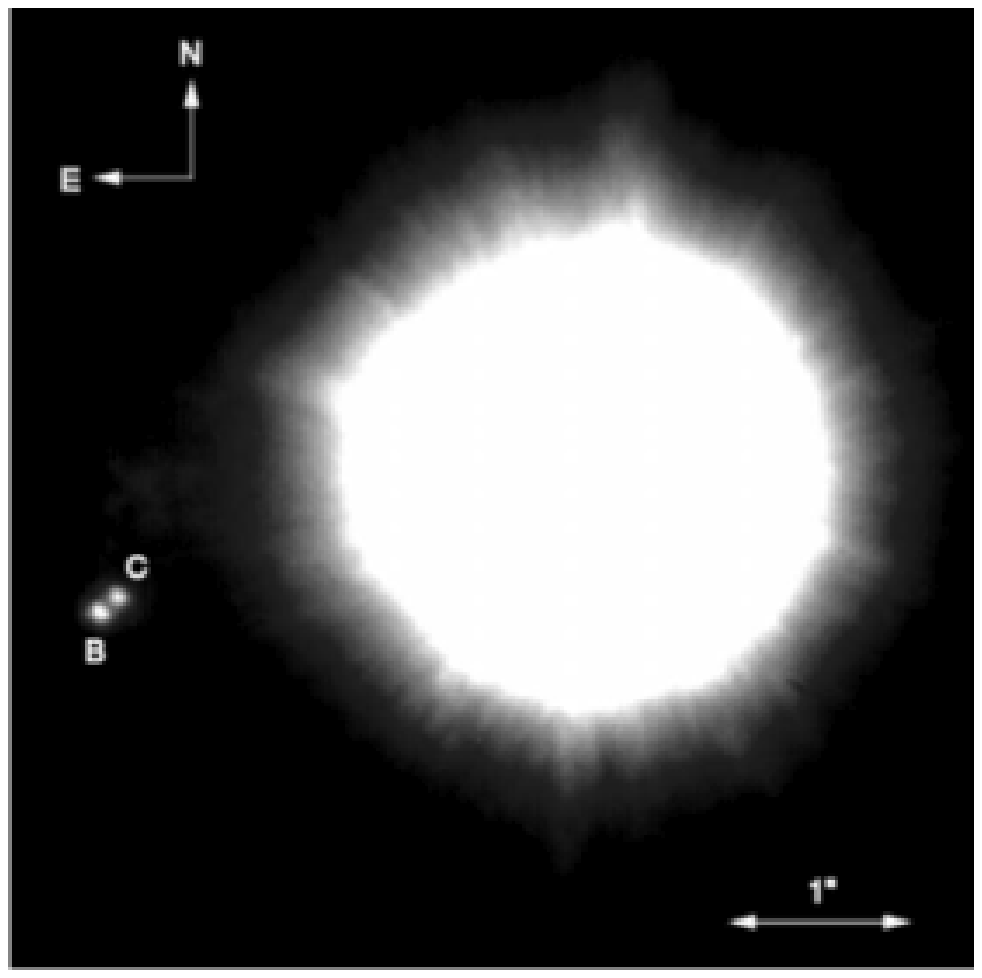}
\hspace{0.5cm}
\includegraphics[width=4.5cm]{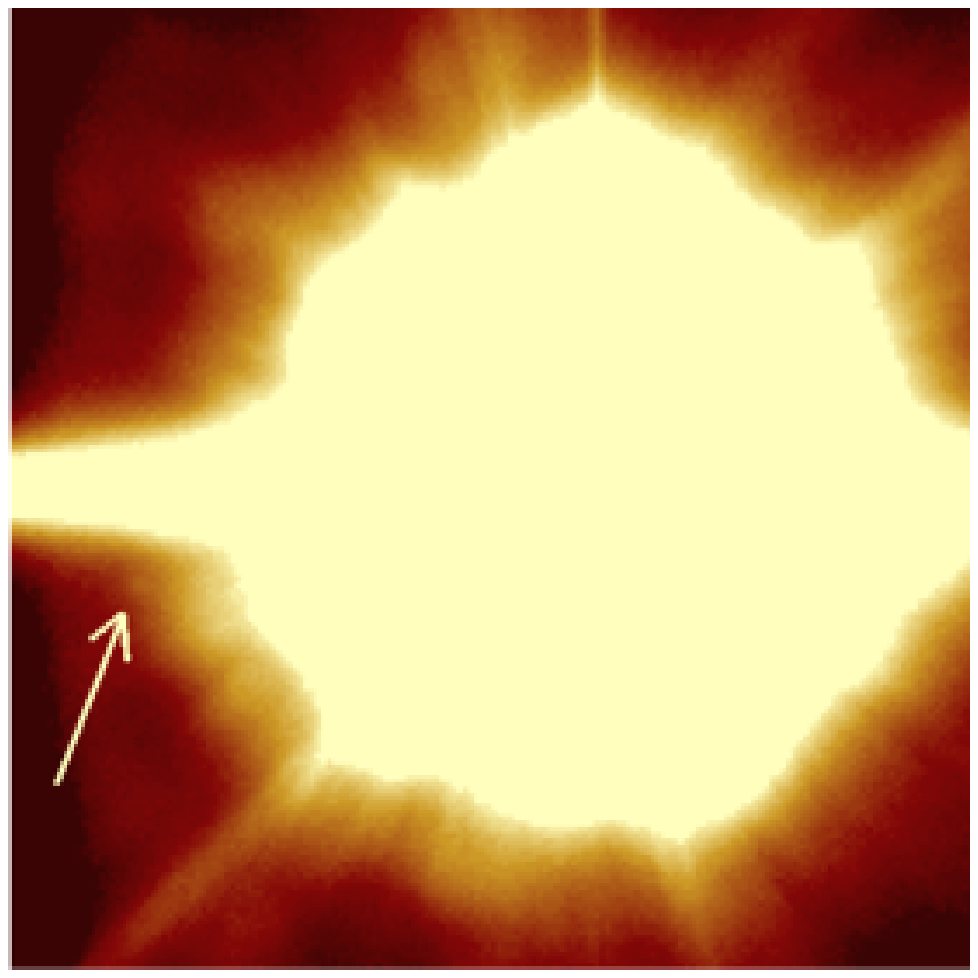}
\hspace{0.5cm}
\includegraphics[width=4.5cm]{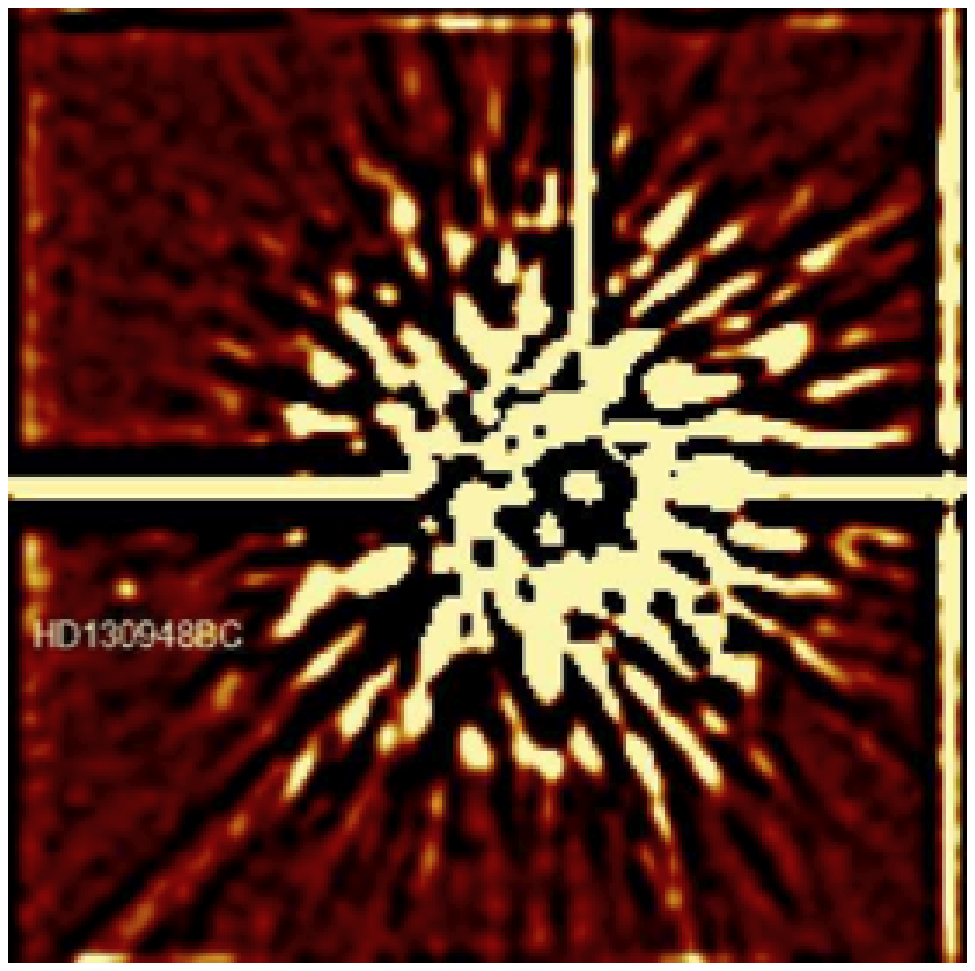}
%\end{minipage}
\caption{{\it Left:} Original detection with the 8-m GEMINI telescope; {\it Center:} Direct lucky image with the 2.5-m NOT; {\it Right:} HD\,130948 system after wavelet image post-processing. In all the images, North is up and East is left. -- Adapted from Labadie et al. (2010)\cite{Labadie2010} .}\label{hd130948}
\end{figure}

\subsection{The case of HD\,130948 system}

In order to evaluate the potential of this approach, we applied the method presented above to the case of HD\,130948 in order to obtain a high contrast image of the system. HD\,130948 is a main sequence G star at $\sim$20\,pc hosting a brown dwarf binary at a separation of $\sim$2.5$^{\prime\prime}$ from the primary. HD\,130948\,BC is a benchmark for the independent determination of dynamical masses of sub-stellar objects, which is a key step for the calibration of the corresponding mass-luminosity relationship. The binary was initially discovered by Potter et al. (2002)\cite{Potter2002} in the near-infrared using the 8-m GEMINI telescope coupled to adaptive optics, and consequently reported as a pair of L2$\pm$2 sub-stellar objects.

\begin{figure}[h]
\centering
%\begin{minipage}[t]{\textwidth}
\includegraphics[width=4.4cm]{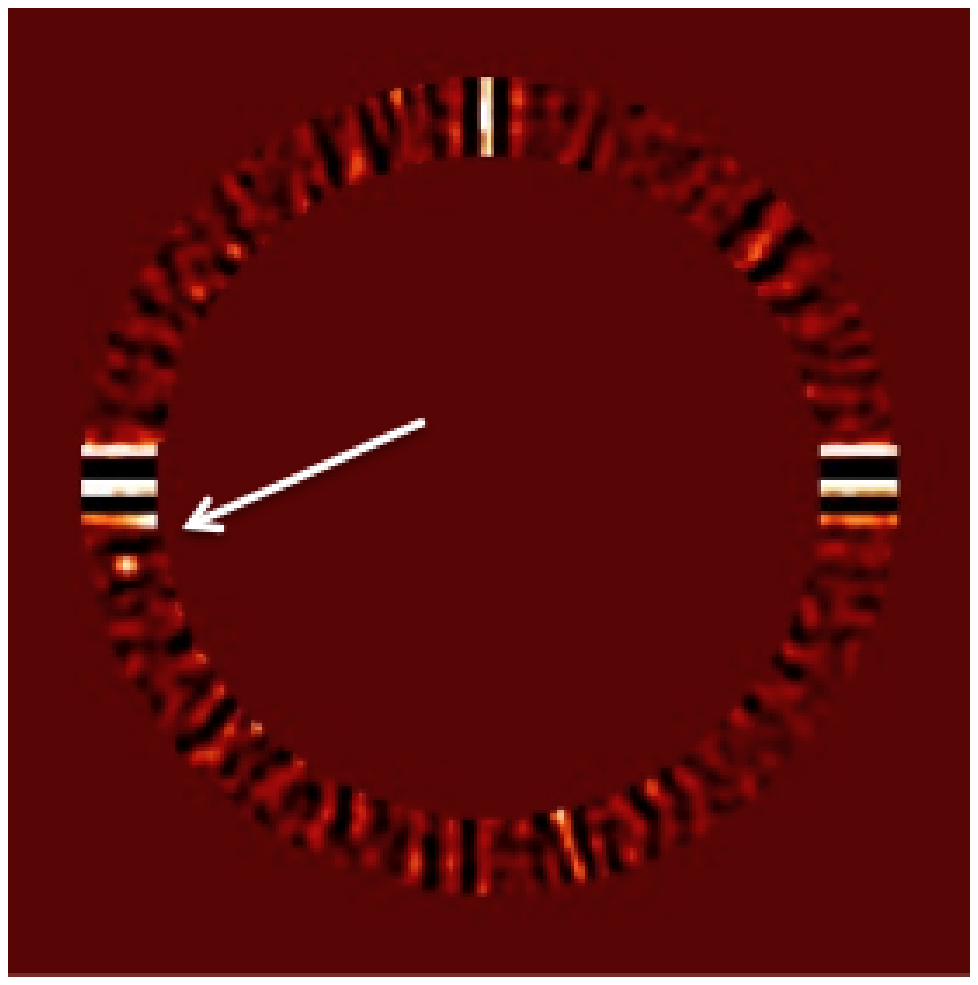}
\hspace{0.5cm}
\includegraphics[width=4.4cm]{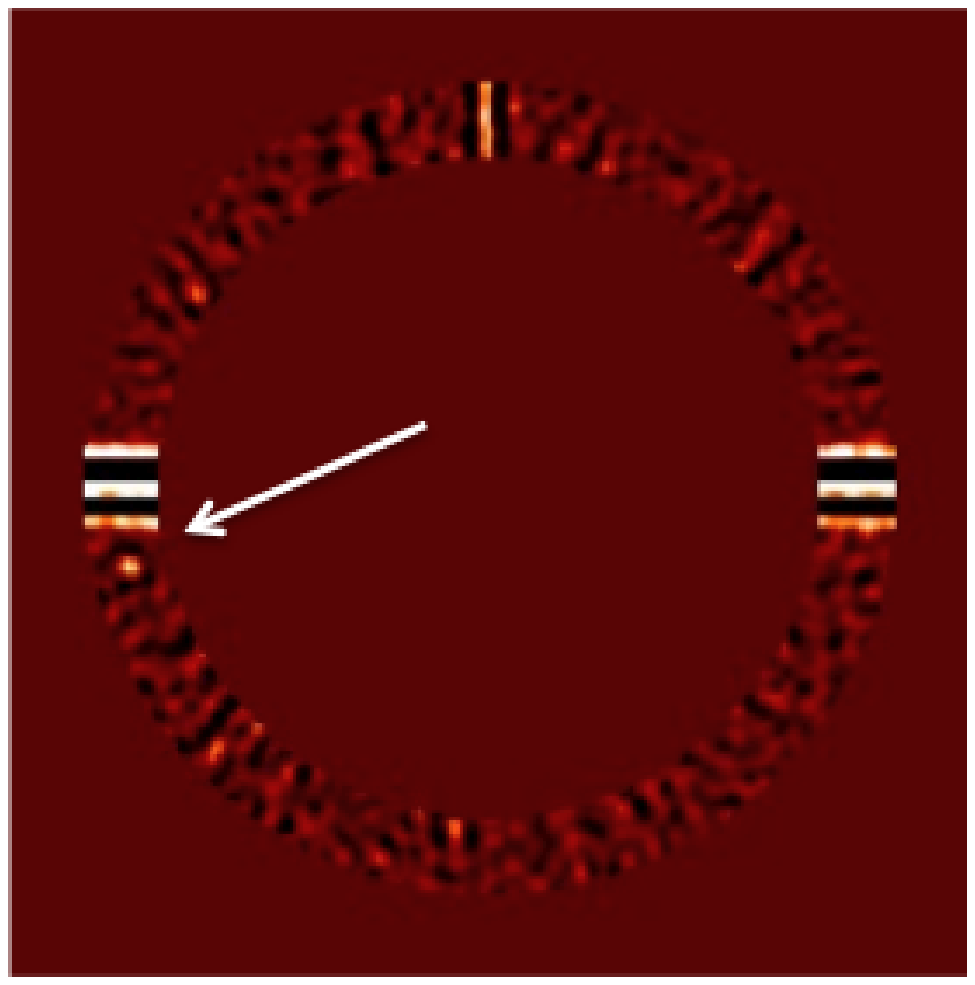}
\hspace{0.5cm}
\includegraphics[width=4.4cm]{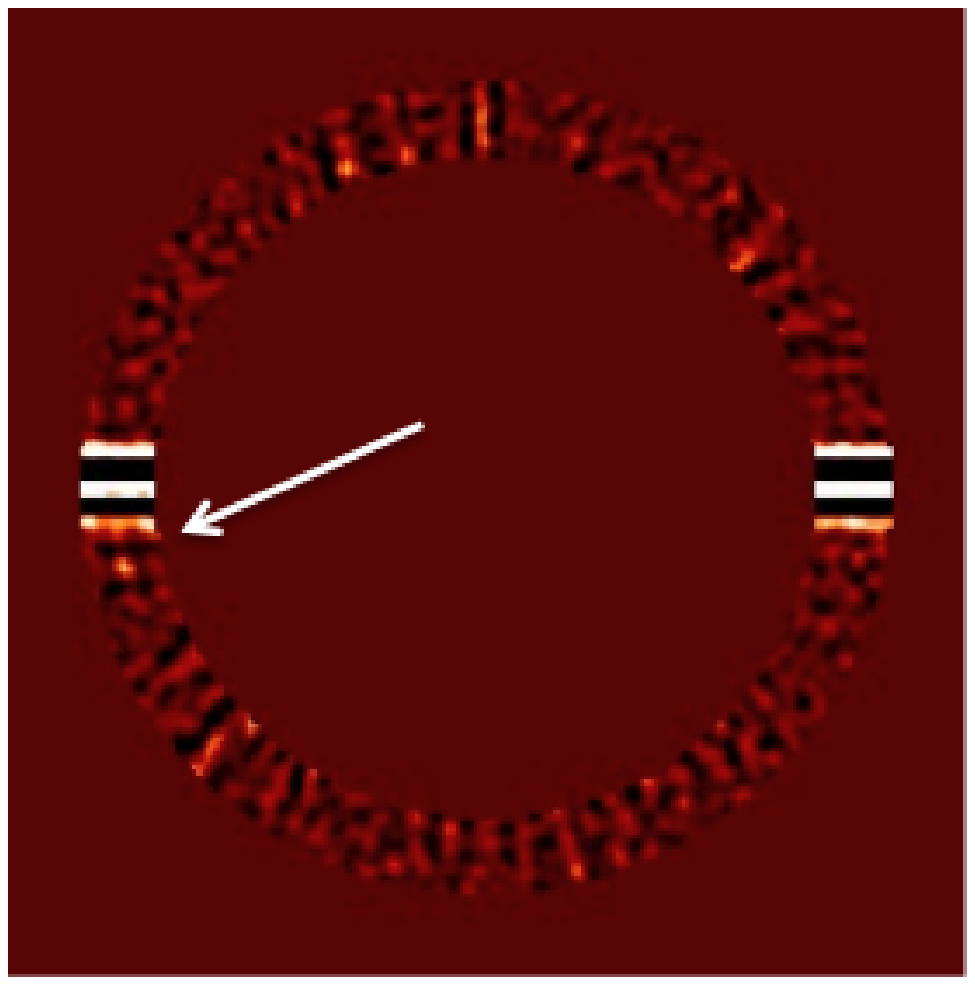}
%\end{minipage}
\caption{Three different time series of the HD\,130948\,BC observations from July 2008, limited to a 2.5$^{\prime\prime}$ annulus. The white arrow indicate the location of the brown dwarf in the image.}\label{time}
\end{figure}

\noindent HD\,130948 was observed with FASTCAM on the 2.5-m Nordic Telescope at two different epochs, in May 2008 and July 2008. We acquired a total number of, respectively, 5$\times$10$^4$ and 10$^5$ images in the $I$ filter. The night conditions were good and excellent for the May and July runs. Fig.~\ref{hd130948} shows respectively the original detection, the direct lucky image and the result after post-processing for the July run. In the last image, we clearly detect a point-like source east from the primary with SNR$\sim$10, which position is consistent with that none for HD\,130948\,BC. In the NOT image, HD\,130948\,BC is not detected as a binary, but as a point source. From a posterior detailed study of the brown dwarf binary orbit\cite{Dupuy2009} , the binary separation at the time of observation was $\sim$35\,mas, beyond the actual diffraction limit of the telescope necessary to resolve the pair.\\
Different tests have been conducted in order to confirm the detection and avoid a false-detection of a speckle. A first test has consisted in splitting the July data in three different time series in order to check if the brown dwarf pair remains detectable at the same location, oppositely to time varying speckle. This is illustrated in the sequence of Fig.~\ref{time}, which presents observations at three different moments in an annulus of appropriate size. At the position of the white arrow is shown the location of HD\,130948\,BC, which is detected in all the three time series, while other bright features are not present in all the three sequences. This confirms that the detected point-source is real. In the last image, the detection of HD\,130948\,BC has less signal-to-noise than in the other two cases: this is because the frames selection for the last time serie resulted in a poorer Strehl ratio -- probably because of degrading weather conditions -- which has spread out the flux of the brown dwarf.

\vspace{0.0cm}

\begin{figure}[h]
\centering
%\begin{minipage}[t]{\textwidth}
\includegraphics[width=5.4cm]{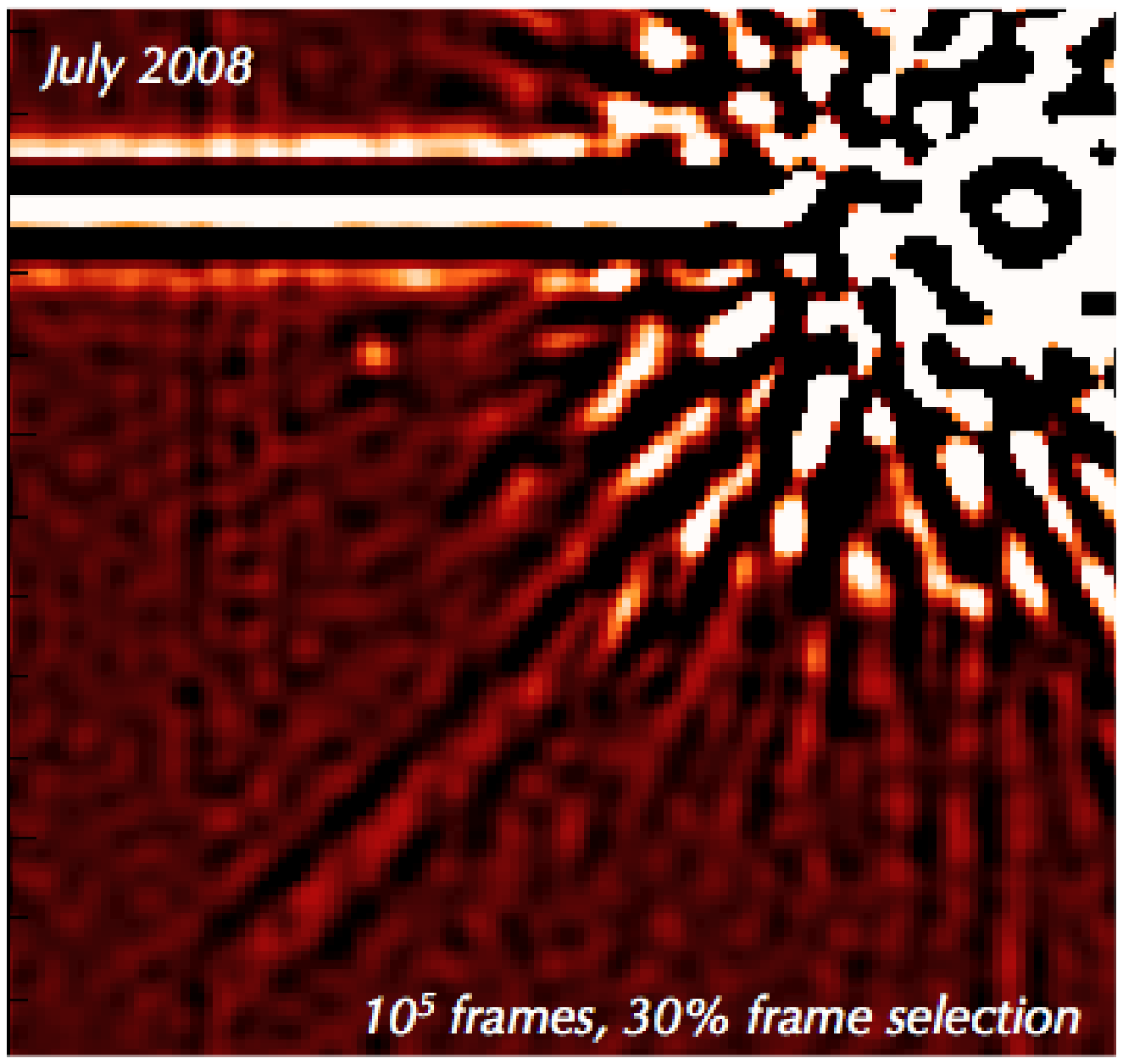}
\hspace{1.5cm}
\includegraphics[width=5.4cm]{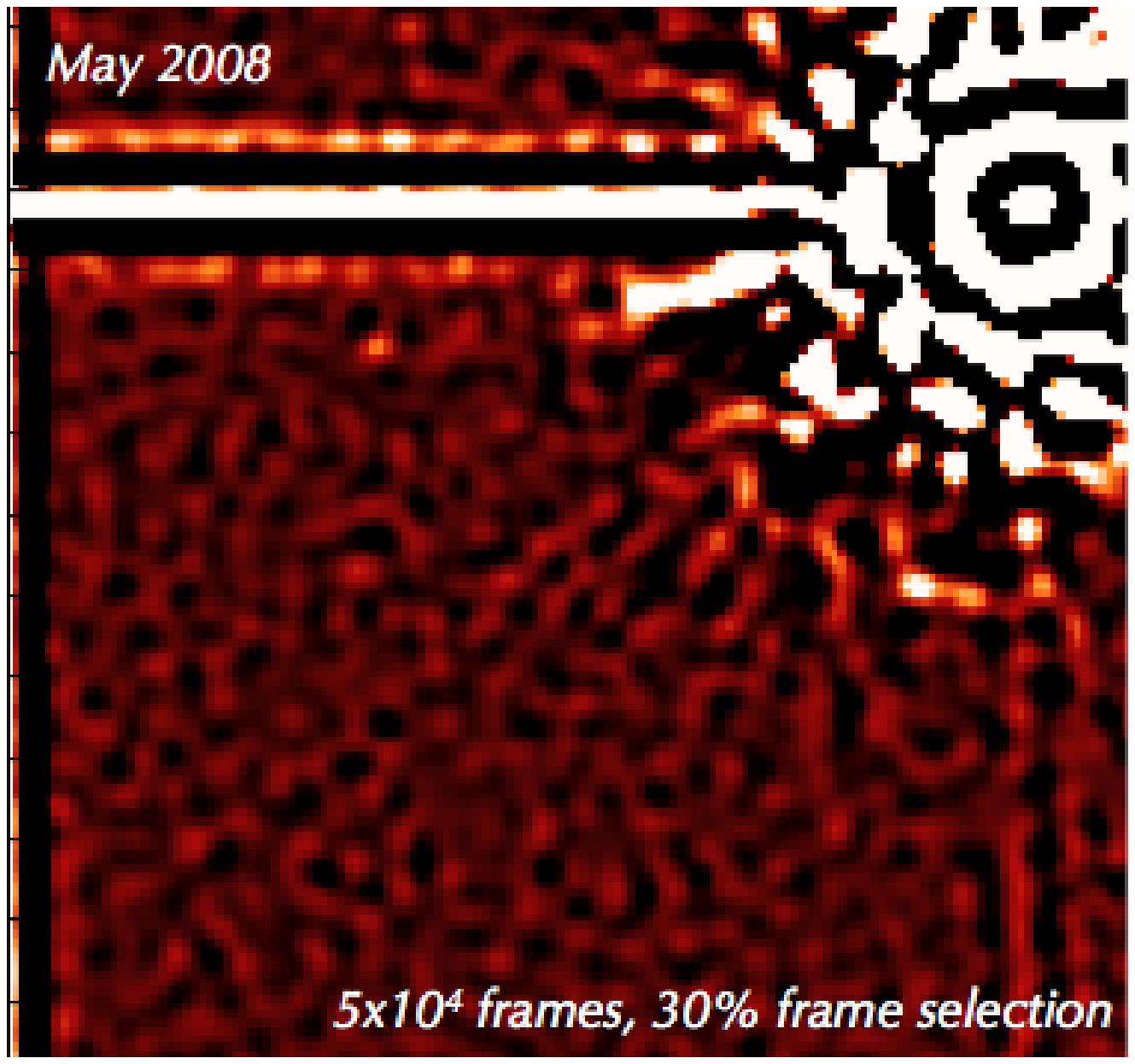}
%\end{minipage}
\caption{Second epoch observation of HD\,130948\,BC}\label{sec_epoch}
\end{figure}

\noindent The ultimate test for the detection of the companion consists in the second epoch observation of May 2008 taken with FASTCAM. The same data reduction procedure is adopted. Despite the lower number of available frames, the detection of HD\,130948\,BC is also confirmed in the May dataset, although the PSF is slightly more spread out than in the July data (cf. Fig.~\ref{sec_epoch}). The two images have been normalized to the peak intensity and can be directly compared.

%\vspace{0.10cm}

\begin{figure}[h]
\centering
\includegraphics[width=5.0cm]{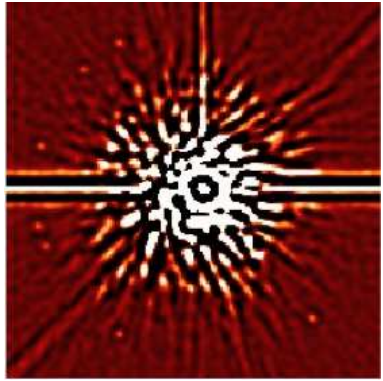}
\caption{Astrometric and photometric check of the effect of the wavelet filtering process}\label{astrometry}
\end{figure}

\noindent In order to evaluate the impact of the wavelet filtering on the photometry and astrometry of the object, we have simulated several companions at a known distance from the primary and with a known magnitude difference (e.g. Fig.~\ref{astrometry}). After having convolved this model with the direct lucky image and applied the filtering process, we measure the position of the photocenter and the peak intensity of the component with respect to the input model. We could verify that the astrometric position was not altered by more than $\sim$0.2 pixels by the filtering process, and that the photometric variation is negligible down to contrasts of $\sim$10$^{-5}$. This result was expected theoretically since the wavelet filtering process is a linear one, which normally keeps the contrast value in the different regions of the image.

\vspace{0.20cm}

\noindent This work on HD\,130948\,BC has permitted us to derive the first measurement of the combined photometry in the $I$-band and examine the astrometric parameters that reveal orbital motion \cite{Labadie2010}. These results suggest that optical speckle imaging has sufficient potential for high dynamic range imaging, even with a 2.5-m telescope, to study faint and close sub-stellar companions in a wavelength regime not efficiently accessible with the current AO systems.

\subsection{Exploring dense cores in globular clusters}

\begin{figure}[b]
\centering
\includegraphics[width=6.0cm]{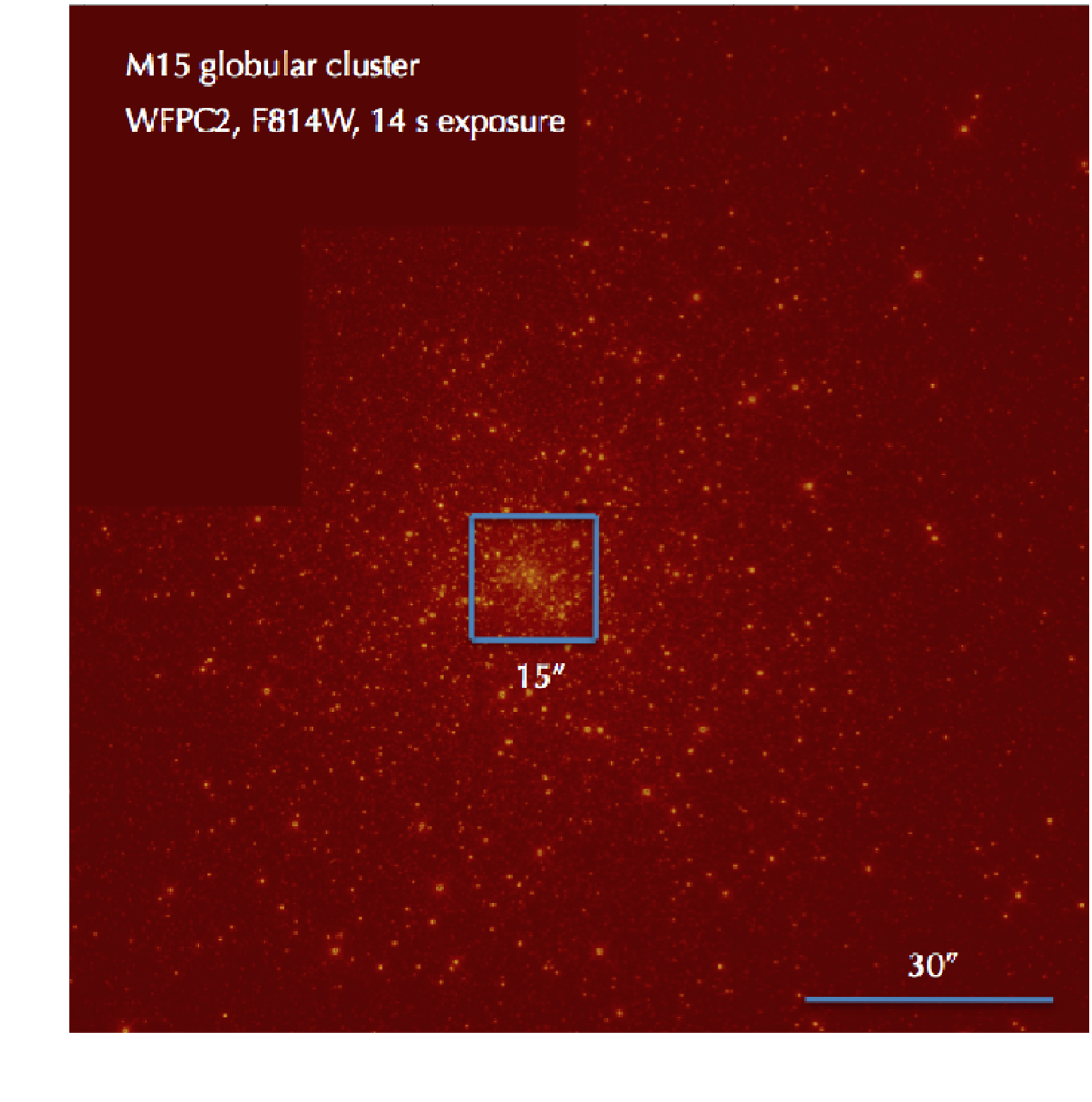}
\hspace{0.5cm}
\includegraphics[width=4.5cm]{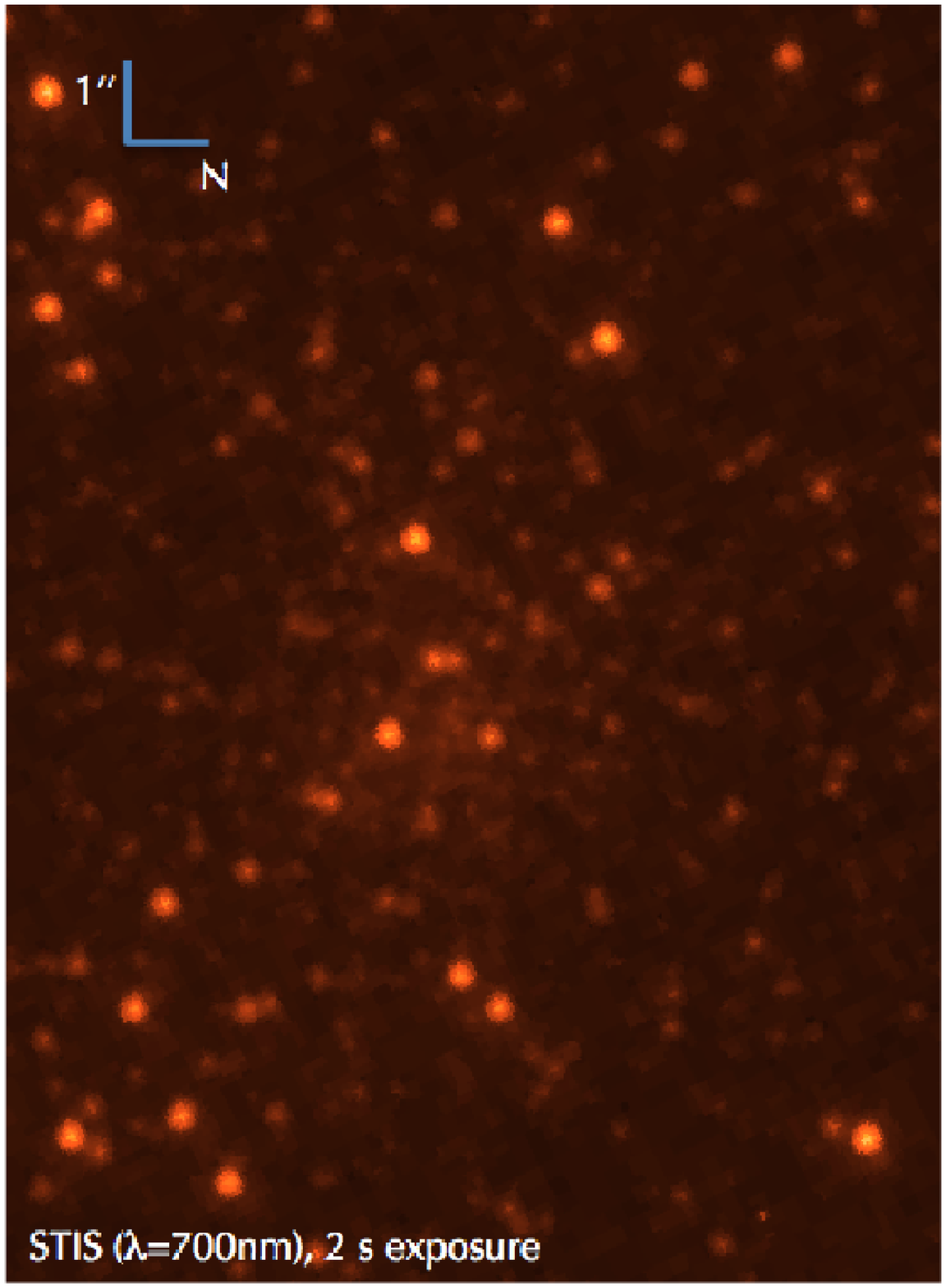}
\caption{HST observations of the M15 cluster and its core at 700 and 800\,nm. See text for details.}\label{hst}
\end{figure}

\begin{figure}[h]
\centering
\includegraphics[width=7.0cm]{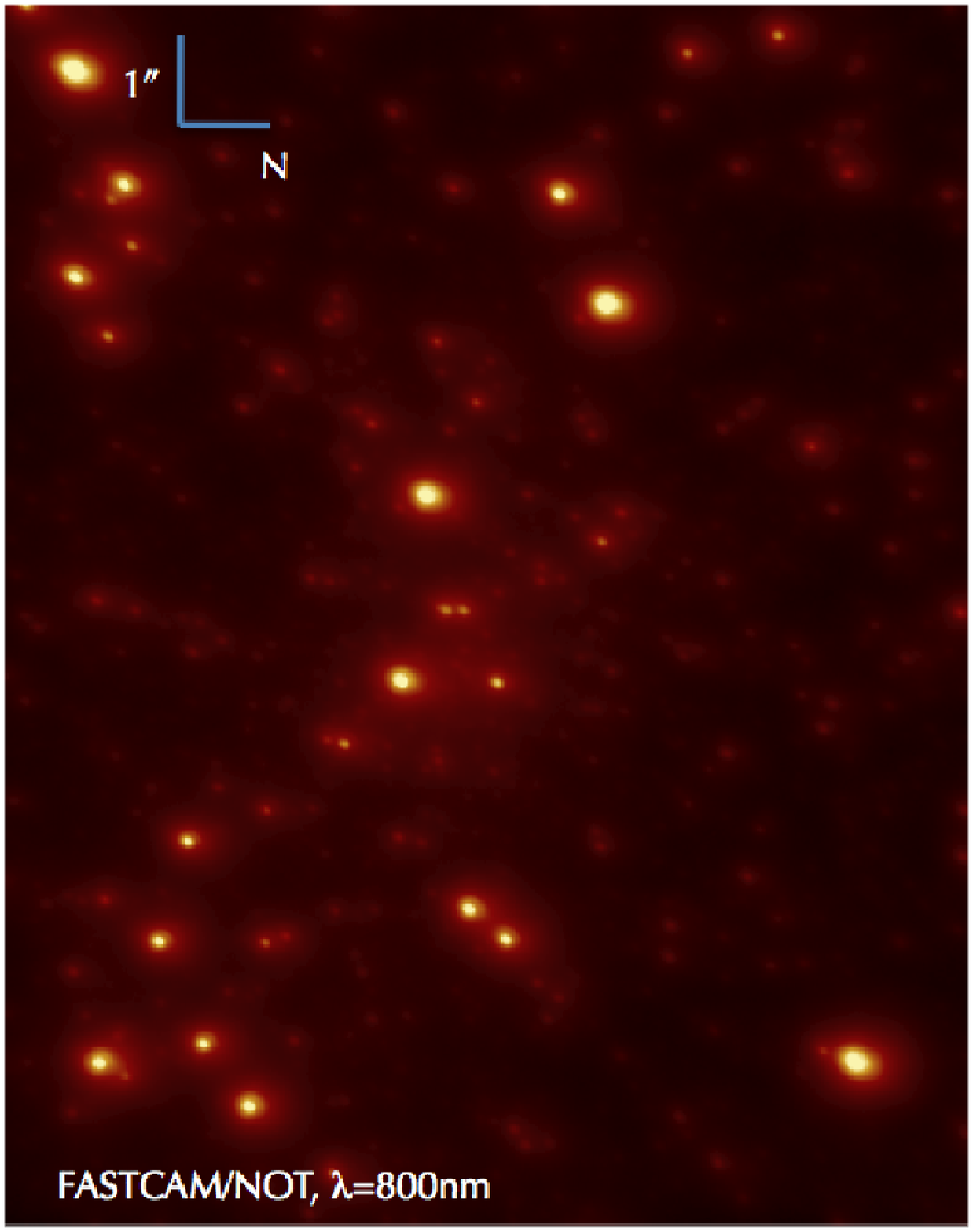}
\hspace{1.0cm}
\includegraphics[width=7.0cm]{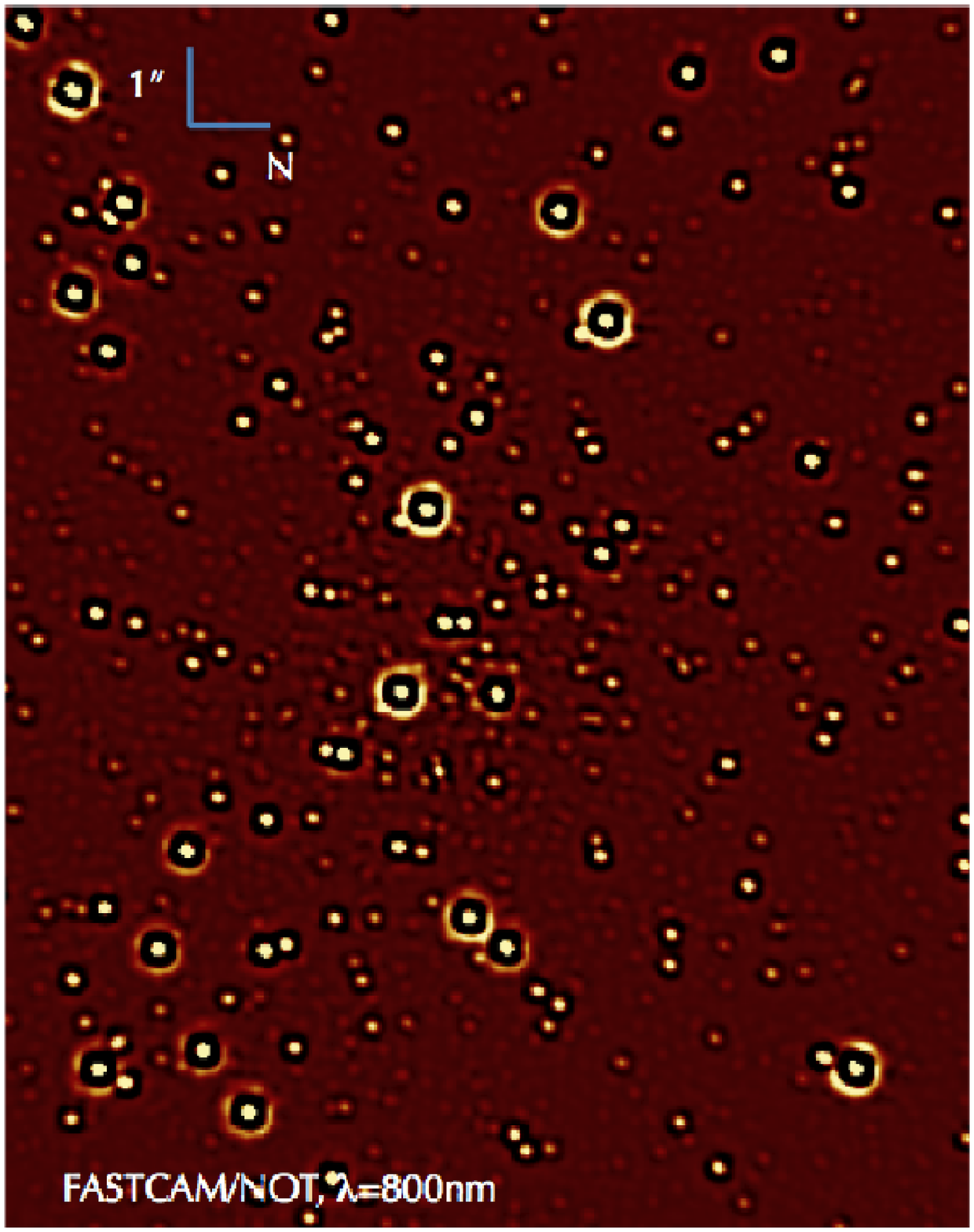}
\caption{FASTCAM observations of the M15 core and corresponding wavelet filtering result. See text for details.}\label{m15-fastcam}
\end{figure}

Another strong application of speckle imaging coupled to image post-processing is high resolution and high contrast imaging in dense cores of globular clusters. These regions are composed of compact groups of thousands of stars in a wide range of magnitudes. The deepest and sharpest observations of these objects are achieved with space-based observatories like HST able to couple high sensitivity and high spatial resolution in the optical. \\
\noindent In Fig.~\ref{hst}, we show two HST optical images of the M\,15 globular cluster. The image on the left side is a wide field-of-view image of the cluster taken in the $I$ band (100\,mas pixel scale) where the dense core is identified in a 15$^{\prime\prime}$$\times$15$^{\prime\prime}$ region. The image on the right is a zoom on this region taken at shorter wavelength and 50\,mas/pix with HST/STIS. In the center of such a region, the aggregation of an elevate number of stars results in a diffuse halo of light able to mask the fainter stars, especially in the red part of the optical spectrum. Although FASTCAM is not comparable to HST in terms of sensitivity, the high spatial resolution and high contrast capabilities of out instrument help us to detect {\it from the ground} a very large fraction of objects in the $I$ band that have optical counterparts detected in $B$ and $V$ bands with HST\cite{Vandermarel2002} .\\
\\
In Fig.~\ref{m15-fastcam} are shown two FASTCAM views of the M15 core, with the same field-of-view as in the right image in Fig.~\ref{hst}. The left view shows the direct lucky image with 7\,\% frame selection (for an effective integration time of $\sim$7\,s; 0.03$^{\prime\prime}$/pixel) taken with the 2.5-m Nordic Telescope. A coarse and direct comparison with the HST image shows the sharpness of the FASTCAM point-spread function, which permits to detect relatively close and faint stars. In the right view of Fig.~\ref{m15-fastcam}, the strong halo-suppression effect of the wavelet filtering applied to the M15 image reveals a high number of faint point sources with contrasts down to $\sim$5$\times$10$^{-4}$\,--\,10$^{-3}$ compared to the brightest stars in the field. The completeness magnitude in the I-band is  $\sim$19. The number of sources detected at 3-$\sigma$ with FASTCAM is above 1000 in a radius of 8$^{\prime\prime}$ from the centre of M15. These source were previously known from HST images at shorter wavelengths.

\vspace{0.20cm}

\section{Conclusions and immediate perspectives}

We have presented in this paper first observational results with the speckle imager FASTCAM in the context of high angular resolution {\it and} high dynamic range imaging with the speckle imaging technique from the ground, in a spectral domain where classic AO is not efficient. We are able to routinely obtain close to diffraction-limited optical images ($I$ band) both with the 2.5-m Nordic Telescope and the 4.2-m William Herschel Telescope on the Observatory of Roque de los Muchachos, in the Canary Islands.\\
Our results demonstrate high sensitivity in crowded fields ($I$$\sim$19), while high contrast of $\sim$12 mag is obtained in the vicinity ($\sim$1--2$^{\prime\prime}$) of bright sources, with a $\sim$2--3 mag gain due to post-processing wavelet filtering of the images. This permitted to detect, without resolving so far, the brown dwarf binary HD\,130948\,BC considered as a benchmark for the independent determination of dynamical masses. The property of {\it unsaturated} PSF in the FASTCAM images permits us, as opposed to classical long-exposure imaging, to carry high accuracy astrometric measurement of brown dwarf binaries and high accuracy relative photometry (see Femenia et al., SPIE Adaptive Optics Systems II (Conference 7736)  proceedings).\\
\\
As the wavelet post-processing filtering can presumably affect the detection of {\it extended} sources, we plan in a near future to explore the option of more classical image reconstruction with speckle imaging, based on speckle holography or phase diversity techniques, in order to improve the dynamic range in the image without filtering out precious spatial frequencies. 
%\vspace{0.20cm}
%\begin{figure}[h]
%\centering
%\includegraphics[width=7.0cm]{m15-holo-1.ps}
%\hspace{1.0cm}
%\includegraphics[width=7.0cm]{m15-holo-2.ps}
%\caption{Speckle holography reconstruction of the M15 core (courtesy R.~Sch\"odel, IAA Granada). See text for details.}\label{m15-holo}
%\end{figure}
%\noindent In Fig.~\ref{m15-holo} are shown preliminary results on how speckle holography can, even at optical wavelengths, help us to reconstruct an impressively sharp image. In the precise case of M15, the reconstruction process is even improved with the use of more than one PSF-reference stars picked up in the field. The brightest stars in the reconstructed field still show  reconstruction artifacts which presence needs further investigation. 
Ideally, these techniques could be applied to the case of imaging of edge-on protoplanetary disks in scattered light at optical wavelenghts, for which nearby ($\sim$5$^{\prime\prime}$) stars could be used as guide stars during the observations and PSF references for image reconstruction purposes.

\acknowledgments     %>>>> equivalent to \section*{ACKNOWLEDGMENTS}       
 
LL and BF are funded by the Spanish MICINN under the Consolider-Ingenio 2010 Program grant CSD2006-00070:First Science with the GTC (www.iac.es/consolider-ingenio-gtc).

%%%%%%%%%%%%%%%%%%%%%%%%%%%%%%%%%%%%%%%%%%%%%%%%%%%%%%%%%%%%%
%%%%% References %%%%%

\bibliography{report}   %>>>> bibliography data in report.bib

\begin{thebibliography}{10}

\bibitem{Bally2000}
{Bally}, J., {O'Dell}, C.~R., and {McCaughrean}, M.~J., ``{Disks, Microjets,
  Windblown Bubbles, and Outflows in the Orion Nebula},'' {\em \aj}~{\bf 119},
  2919--2959 (June 2000).

\bibitem{Bouy2004}
{Bouy}, H., {Duch{\^e}ne}, G., {K{\"o}hler}, R., {Brandner}, W., {Bouvier}, J.,
  {Mart{\'{\i}}n}, E.~L., {Ghez}, A., {Delfosse}, X., {Forveille}, T.,
  {Allard}, F., {Baraffe}, I., {Basri}, G., {Close}, L., and {McCabe}, C.~E.,
  ``{First determination of the dynamical mass of a binary L dwarf},'' {\em
  \aap}~{\bf 423},  341--352 (Aug. 2004).

\bibitem{Zapatero2004}
{Zapatero Osorio}, M.~R., {Lane}, B.~F., {Pavlenko}, Y., {Mart{\'{\i}}n},
  E.~L., {Britton}, M., and {Kulkarni}, S.~R., ``{Dynamical Masses of the
  Binary Brown Dwarf GJ 569 Bab},'' {\em \apj}~{\bf 615},  958--971 (Nov.
  2004).

\bibitem{Mackay2004}
{Mackay}, C.~D., {Baldwin}, J., {Law}, N., and {Warner}, P., ``{High-resolution
  imaging in the visible from the ground without adaptive optics: new
  techniques and results},'' in [{\em Society of Photo-Optical Instrumentation
  Engineers (SPIE) Conference Series}{\nolinebreak\hspace{0.1em}]},
  {A.~F.~M.~Moorwood \& M.~Iye}, ed., {\em Presented at the Society of
  Photo-Optical Instrumentation Engineers (SPIE) Conference} {\bf 5492},
  128--135 (Sept. 2004).

\bibitem{Law2006}
{Law}, N.~M., {Mackay}, C.~D., and {Baldwin}, J.~E., ``{Lucky imaging: high
  angular resolution imaging in the visible from the ground},'' {\em \aap}~{\bf
  446},  739--745 (Feb. 2006).

\bibitem{Oscoz2008}
{Oscoz}, A., {Rebolo}, R., {L{\'o}pez}, R., {P{\'e}rez-Garrido}, A.,
  {P{\'e}rez}, J.~A., {Hildebrandt}, S., {Rodr{\'{\i}}guez}, L.~F., {Piqueras},
  J.~J., {Vill{\'o}}, I., {Gonz{\'a}lez}, J.~M., {Barrena}, R., {G{\'o}mez},
  G., {Garc{\'{\i}}a}, A., {Monta{\~n}{\'e}s}, P., {Rosenberg}, A., {Cadavid},
  E., {Calcines}, A., {D{\'{\i}}az-S{\'a}nchez}, A., {Kohley}, R.,
  {Mart{\'{\i}}n}, Y., {Pe{\~n}ate}, J., and {S{\'a}nchez}, V., ``{FastCam: a
  new lucky imaging instrument for medium-sized telescopes},'' in [{\em Society
  of Photo-Optical Instrumentation Engineers (SPIE) Conference
  Series}{\nolinebreak\hspace{0.1em}]},  {\em Presented at the Society of
  Photo-Optical Instrumentation Engineers (SPIE) Conference} {\bf 7014} (Aug.
  2008).

\bibitem{Labeyrie1970}
{Labeyrie}, A., ``{Attainment of Diffraction Limited Resolution in Large
  Telescopes by Fourier Analysing Speckle Patterns in Star Images},'' {\em
  \aap}~{\bf 6},  85--+ (May 1970).

\bibitem{Weigelt1977}
{Weigelt}, G.~P., ``{Modified astronomical speckle interferometry 'speckle
  masking'},'' {\em Optics Communications}~{\bf 21},  55--59 (Apr. 1977).

\bibitem{Correia2006}
{Correia}, S., {Zinnecker}, H., {Ratzka}, T., and {Sterzik}, M.~F., ``{A
  VLT/NACO survey for triple and quadruple systems among visual pre-main
  sequence binaries},'' {\em \aap}~{\bf 459},  909--926 (Dec. 2006).

\bibitem{Potter2002}
{Potter}, D., {Mart{\'{\i}}n}, E.~L., {Cushing}, M.~C., {Baudoz}, P.,
  {Brandner}, W., {Guyon}, O., and {Neuh{\"a}user}, R., ``{Hokupa'a-Gemini
  Discovery of Two Ultracool Companions to the Young Star HD 130948},'' {\em
  \apjl}~{\bf 567},  L133--L136 (Mar. 2002).

\bibitem{Labadie2010}
{Labadie}, L., {Rebolo}, R., {Vill\`o}, I., {Perez-Prieto}, J.~A.~P.,
  {Perez-Garrido}, A., {Hildebrandt}, S.~R., {Femen\`ia}, B., {Oscoz}, A., and
  {Lopez}, R., ``{High contrast optical imaging of companions: the case of the
  brown dwarf binary HD130948 BC},'' {\em submitted to \aa} .

\bibitem{Dupuy2009}
{Dupuy}, T.~J., {Liu}, M.~C., and {Ireland}, M.~J., ``{Dynamical Mass of the
  Substellar Benchmark Binary HD 130948BC},'' {\em \apj}~{\bf 692},  729--752
  (Feb. 2009).

\bibitem{Vandermarel2002}
{van der Marel}, R.~P., {Gerssen}, J., {Guhathakurta}, P., {Peterson}, R.~C.,
  and {Gebhardt}, K., ``{Hubble Space Telescope Evidence for an
  Intermediate-Mass Black Hole in the Globular Cluster M15. I. STIS
  Spectroscopy and WFPC2 Photometry},'' {\em \aj}~{\bf 124},  3255--3269 (Dec.
  2002).

\end{thebibliography}
\bibliographystyle{spiebib}   %>>>> makes bibtex use spiebib.bst

\end{document}